\documentclass[twocolumn]{aastex631}

\usepackage{xspace}
\newcommand{\pdsb}{PDS\,70\,b\xspace}
\newcommand{\pdsc}{PDS\,70\,c\xspace}
\newcommand{\pdsbc}{PDS\,70\,b and c\xspace}

\newcommand{\halpha}{H\ensuremath{\alpha}\xspace}

\newcommand{\MJ}{M_{\mathrm{Jup}}}

\newcommand{\fluxcflamI}{\ensuremath{3.04}\xspace}
\newcommand{\fluxcepsII}{\ensuremath{2.41\pm0.44}\xspace}
\newcommand{\fluxcflamII}{\ensuremath{7.3\pm1.3}\xspace}

\begin{document}

\title{Evidence for Variable Accretion onto \pdsc and Implications for Protoplanet Detections}  
\shorttitle{PDS\,70\,\lowercase{c} is Variable.}

\correspondingauthor{Yifan Zhou}
\email{yzhou@virginia.edu}

\author[0000-0003-2969-6040]{Yifan Zhou}
\affiliation{University of Virginia, 530 McCormick Rd. Charlottesville, VA 22904, USA}

\author[0000-0003-2649-2288]{Brendan P. Bowler}
\affiliation{Department of Astronomy, University of Texas at Austin, 2515 Speedway Stop C1400, TX 78712, USA}

\author[0000-0002-1838-4757]{Aniket Sanghi}
\altaffiliation{NSF Graduate Research Fellow.}
\affiliation{Cahill Center for Astronomy and Astrophysics, California Institute of Technology, 1200 E. California Boulevard, MC 249-17, Pasadena, CA 91125, USA}	

\author[0000-0002-2919-7500]{Gabriel-Dominique Marleau}
\affiliation{Max-Planck-Institut f\"ur Astronomie, K\"onigstuhl 17,
69117 Heidelberg, Germany}
\affiliation{Fakult\"at f\"ur Physik, Universit\"at Duisburg-Essen, Lotharstra\ss{}e 1,
47057 Duisburg, Germany}
\affiliation{Physikalisches Institut, Universit\"{a}t Bern, Gesellschaftsstr.~6, 3012 Bern, Switzerland}

\author[0000-0003-3882-3945]{Shinsuke Takasao}
\affiliation{Department of Earth and Space Science, Graduate School of Science, Osaka University, Toyonaka, Osaka 560-0043, Japan}

\author[0000-0003-0568-9225]{Yuhiko Aoyama}
\affiliation{School of Physics and Astronomy, Sun Yat-sen University, Zhuhai 519082, P.R. China}	

\author[0000-0002-9017-3663]{Yasuhiro Hasegawa}
\affiliation{Jet Propulsion Laboratory, California Institute of Technology, Pasadena, CA 91109, USA}

\author[0000-0003-4507-1710]{Thanawuth Thanathibodee}
\affiliation{Department of Physics, Faculty of Science, Chulalongkorn University, 254 Phayathai Road, Pathumwan, Bangkok 10330, Thailand}	

\author[0000-0002-6879-3030]{Taichi Uyama}
\affiliation{Department of Physics \& Astronomy, California State University Northridge, Live Oak Hall, Room 1128
18111 Nordhoff Street
Northridge, CA 91330, USA}	

\author[0000-0002-3053-3575]{Jun Hashimoto}
\affil{Astrobiology Center, National Institutes of Natural Sciences, 2-21-1 Osawa, Mitaka, Tokyo 181-8588, Japan}
\affil{Subaru Telescope, National Astronomical Observatory of Japan, Mitaka, Tokyo 181-8588, Japan}
\affil{Department of Astronomy, School of Science, Graduate University for Advanced Studies (SOKENDAI), Mitaka, Tokyo 181-8588, Japan}	

\author[0000-0002-4309-6343]{Kevin Wagner}
\affiliation{Steward Observatory, University of Arizona, 933 N. Cherry Ave, Tucson, AZ 89712, USA}

\author[0000-0002-3950-5386]{Nuria Calvet}
\affiliation{Department of Astronomy, University of Michigan, 323 West Hall, 1085 South University Avenue, Ann Arbor, MI 48109-1107, USA}

\author[0009-0007-9582-553X]{Dorian Demars}
\affiliation{Universit\'e Grenoble Alpes, CNRS, IPAG, F-38000 Grenoble, France}

\author[0000-0002-4392-1446]{Ya-Lin Wu}
\affiliation{Department of Physics, National Taiwan Normal University, 88 Sec.~4, Tingzhou Road, Taipei 11677, Taiwan}

\author[0000-0003-2646-3727]{Lauren I. Biddle}
\affiliation{Department of Astronomy, University of Texas at Austin, 2515 Speedway Stop C1400, TX 78712, USA}

\author[0000-0001-5130-9153]{Sebastiaan Y. Haffert}
\affiliation{Leiden Observatory, Leiden University, Einsteinweg 55, Leiden 2333CC, The Netherlands}

\author[0000-0002-6076-5967]{Marta L. Bryan}
\affiliation{Depart of Astronomy \& Astrophysics, University of Toronto, 50 St George St, Toronto, ON M5S 3H4, Canada}



\begin{abstract}
Understanding the processes of planet formation and accretion in young systems is essential to unraveling the initial conditions of planetary systems. The PDS\,70 system, which hosts two directly imaged protoplanets, provides a unique laboratory for studying these phenomena, particularly through H$\alpha$ emission, a commonly used accretion tracer. We present multi-epoch observations and examine the variability in accretion signatures within this system, focusing on \pdsbc. Using Hubble Space Telescope narrowband H$\alpha$ imaging from 2020 and 2024, we achieve high signal-to-noise detections of these planets and reveal significant changes in H$\alpha$ flux. For PDS 70 c, the H$\alpha$ flux more than doubled between 2020 and 2024. The trend is consistent with the one identified in recently published MagAO-X data, further confirming that \pdsc has become significantly brighter in H$\alpha$ between 2023 March and 2024 May. The observed variability suggests dynamic accretion processes, possibly modulated by circumplanetary disk properties or transient accretion bursts. High-amplitude variability in \pdsc motivates simultaneous monitoring of multiple accretion tracers to probe the mechanisms of mass growth of gas giant planets. We quantify the impact of variability on the detectability of protoplanets in imaging surveys and emphasize the need for continued and regular monitoring to accurately assess the occurrence and characteristics of young, forming planets. 
\end{abstract}

\keywords{}


\section{Introduction} \label{sec:intro}

The PDS\,70 system is among the most thoroughly studied sites of ongoing planet formation. This solar-mass star harbors two gas giants, \pdsbc, located in the disk cavity \citep{Keppler2018, Haffert2019}. The presence of a third planet has been suggested based on observations from VLT/SPHERE, and more recently, JWST/NIRCam \citep{Mesa2019,Christiaens2024}. There is strong evidence suggesting ongoing mass accretion onto \pdsbc \citep[e.g.,][]{Wagner2018,Haffert2019}, most likely through their circumplanetary disks \citep{Isella2019, Benisty2021}. Additional signs suggest that gas and dust from the outer circumstellar disk are flowing into the disk cavity, fuelling the growth of the two planets and replenishing the inner stellar disk \citep{Perotti2023,Christiaens2024,Gaidos2024}.

Accretion onto the PDS\,70 planets is directly probed by the ultraviolet excess flux tracing the continuum of accretion shock emission \citep{Zhou2021}. However, the hydrogen recombination lines, H$\alpha$ in particular, are more regularly used in the investigation of the mass accretion rates of \pdsb and c \citep[e.g.,][]{Wagner2018,Haffert2019,Hashimoto2020,Zhou2021}, due to the more favorable observational circumstances in the optical bands for detecting H$\alpha$ than the Balmer continuum at these planets. H$\alpha$ emission may originate either from the accretion shock \citep{Aoyama2018,Aoyama2019} or from pre-shock gas in a magnetospheric accretion scenario similar to that of classical T Tauri stars \citep[e.g.,][]{Gullbring1998, Thanathibodee2019}. Depending on the assumed accretion models or scaling relations \citep[e.g.,][]{Natta2004,Alcala2017,Aoyama2021}, the estimated mass accretion rates of \pdsbc range from $10^{-8}$ to $10^{-6}$\,$\MJ\,\mathrm{yr}^{-1}$ \citep[e.g.,][]{Wagner2018,Haffert2019, Hashimoto2020}. In either scenario, most of the mass in these gas giant planets is expected to have already been accreted, since otherwise the current accretion rates could not support the formation of the two super Jupiters within the age of the system (5.4\,Myr, \citealt{Mueller2018}).

The strong \halpha excess emission of the planets, when paired with the weak \halpha{} excess of the star, mitigates the brightness contrast challenge in direct imaging detections \citep[e.g.,][]{Close2014,Zhou2014}. For \pdsbc, the planet-to-star relative brightness in \halpha{} is greater than those in the $J/H/K$ bands and are comparable to those in the $L'$ band \citep{Haffert2019}. Motivated by these findings, multiple H$\alpha$ high-contrast imaging surveys have been conducted to search for accreting (proto-)planets but have largely yielded non-detections of young giant planets \citep{Cugno2019,Zurlo2020,Xie2020}. Interesting \mbox{(proto-)}planetary candidates detected by infrared imaging (e.g.,  \citealt{Sallum2015, Currie2022,Hammond2023,Wagner2023}) or disk kinematic (e.g., \citealt{Teague2018,Pinte2019,Pinte2020}) do not resemble \pdsb or c, i.e., these candidates are either not point sources or lack evidence of accretion \citep[e.g.,][]{Zhou2022}. The low yields defy the expectation \citep{Plunkett2025}. Proposed explanations of the low detection rate of accreting planets include stringent requirements for circumplanetary disk formation \citep[e.g.,][]{Sagynbayeva2024}, extinction due to disk material \citep[e.g.,][]{Hasegawa2024}, rarity of widely separated planets in general \citep[e.g.,][]{Bowler2016,Nielsen2019}, and variability of the \halpha{} emission \citep{Brittain2020}.

Variability is a hallmark of accreting young stars and has been recently detected in accreting substellar companions \citep{Demars2023}. In accreting stars, variability manifests as rotation modulations caused by hot spots and stochastic variations associated with unstable mass flow \citep{Costigan2014}. Instabilities in the accretion process extend to the planetary regime, leading to predictions of variable accretion rate and luminosity in planetary accretion models \citep{Szulagyi2020, Takasao2021}.

Are \pdsbc rare or even unique outcomes of planet formation, or are these planets caught at a fortunate time when the accretion excess emission is especially high? Answers to these questions, which are critical for inferring the intrinsic population of forming gas giants, require monitoring the variability of these planets. Temporal variability in accreting planets also holds the key to the geometry of the accretion and mechanisms of mass growth.

Recently, \citet{Close2025} presented the first definitive evidence of \halpha{} variability in both \pdsbc{} based on three years of ground-based observations with MagAO-X. The \halpha{}  luminosity of \pdsb decreased by a factor of 4.6 between March 2022 and March 2023 and then slightly rose by a factor of 1.6  between March 2023 and March 2024. In contrast, \pdsc's \halpha{} luminosity increased by $2.3\times$ between 2023 and 2024.

In this Letter, we present new observations of the PDS 70 system obtained with Hubble Space Telescope (HST) Wide Field Camera 3 (WFC3) with the F656N narrowband H$\alpha$ filter (Section~\ref{sec:obs}). By combining these images with archival HST H$\alpha$ observations, we construct a dataset with the longest total integration time and temporal coverage of PDS\,70 at these wavelengths to date. This enables high signal-to-noise recoveries of \pdsbc and reveals the transition disk in scattered light. Our analysis provides clear evidence of variable accretion onto the planets (Section~\ref{sec:results}) and places these findings within the broader context of protoplanet detection and planetary accretion processes (Section~\ref{sec:discussion}).

\section{Observations and Data Reduction}
\label{sec:obs}

\begin{figure*}[!t]
    \centering
    \includegraphics[width=\textwidth]{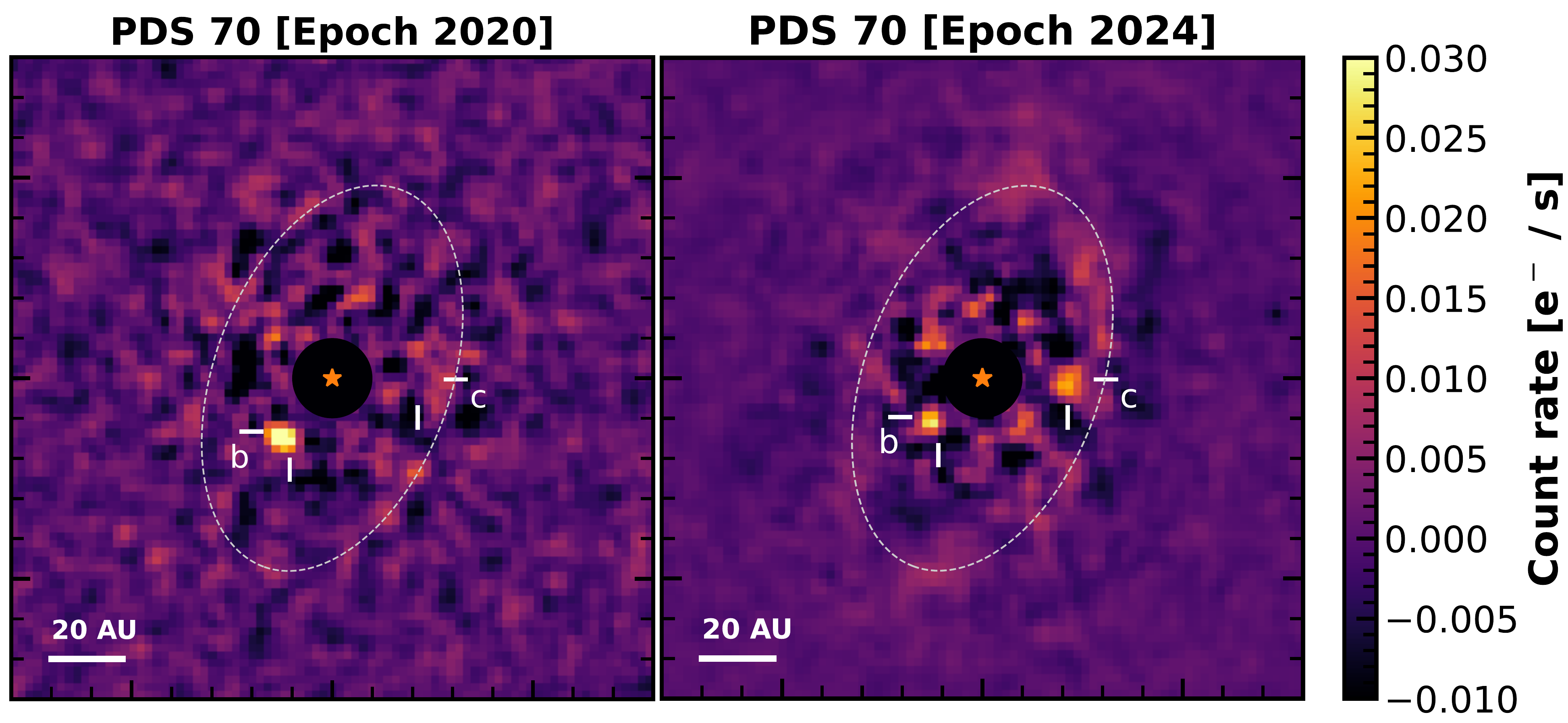}
    \caption{Primary-subtracted images of PDS\,70 in 2020 (left) and 2024 (right). North is up and East is to the left. Color scales of the two panels are identical. The expected positions of \pdsbc are marked. The white dashed line marks the inner edge of PDS\,70's outer disk in infrared scattered light. \pdsb is detected in both epochs and demonstrates a significant orbital motion. \pdsc is only detected in Epoch 2024. The apparent brightness variations in \pdsbc{} do not accurately represent the planets' intrinsic variability, as high-contrast imaging post-processing introduces flux loss. For a comprehensive analysis of the evidence for variability, refer to Section~\ref{sec:evidence}.}\label{fig:images}
\end{figure*}

We observed PDS 70 using the Hubble Space Telescope Wide Field Camera 3 (HST/WFC3) UVIS2 channel. Images were captured with the narrowband F656N filter ($\lambda_\mathrm{cen}=6561.6$\,\AA, $\mathrm{width_{eff}}=17.65$\,\AA) to measure the H$\alpha$ emission from the planetary system. The observations consisted of four visit-sets, starting on UT 2024-04-08, 2024-04-10, 2024-04-12, and 2024-05-15, respectively. All images were collected prior to the telescope's transition to reduced gyro mode in June 2024, and therefore, are not affected by the scheduling and roll angle flexibility of the telescope. The first visit-set included six consecutive orbits, while the remaining three visit-sets comprised two orbits. In three orbits (Orbits 4 and 5 in Visit-set 1 and Orbit 2 in Visit-set 2), the telescope failed to acquire the targets, and the corresponding data were discarded\footnote{A repeat of these observations is scheduled in Spring 2025.}. In total, nine orbits of successful observations have been completed, yielding 216 raw frames. Details of the observations are summarized in the Appendix~\ref{sec:appendix}.

The observing strategy followed the methods described in \citet{Zhou2021}. Specific techniques included four-point half-pixel dithering, Nyquist-sampled image reconstruction using Fourier interlacing \citep{Lauer1999}, and multi-roll angular differential imaging \citep[ADI, e.g.,][]{Schneider1999,Liu2004,Marois2006}. The nine orbits of observations were carried out using six distinctive roll angles to facilitate ADI. The total integration time of 12,600 seconds in the F656N band was nearly four times the F656N data used in \citet{Zhou2021}, significantly improving the photometric sensitivity and precision. 

Data reduction started with the \texttt{flc} files downloaded from the (Mikulski Archive for Space Telescopes) MAST archive. Bad pixels and cosmic rays were initially flagged using the data quality array. Additional visual examination further identified spurious pixels that were not detected by the \texttt{Calwf3} pipeline. These pixels were replaced by bi-linear interpolation of the neighboring pixels. Then, every set of four dithered images are combined into one following the procedures described in \citet{Lauer1999} and \citet{Zhou2021}. The reconstructed image has a pixel size of 0.020 arcsec. Finally, geometric correction was conducted using the solution provided by \citet{Bellini2011}.

We used the principal component analysis-based method implemented by the \texttt{pyKLIP} package to model and subtract the point spread function (PSF) of the host star \citep{Soummer2012}. The ADI cube combines data taken in this observation (referred to as Epoch 2024) and data published in \citet{Zhou2021} (referred to as Epoch 2020). To aid in PSF modeling, an external reference star image library is formed using archival high-contrast images of several transition disk hosts that were observed using the same instrument set-ups obtained in program GO-16651. For the PSF modeling strategy, we have experimented with ADI, reference star differential imaging (RDI), and the combination of ADI and RDI (ADI + RDI). The ADI + RDI strategy was adopted because it delivered the deepest contrast.  The \texttt{pyKLIP} control variables affecting the performance of PSF subtraction (b: \texttt{movement}=3, \texttt{subsection}=3, \texttt{annuli}=15, \texttt{numbasis}=30; c: \texttt{movement}=3, \texttt{subsection}=1, \texttt{annuli}=15, \texttt{numbasis}=15) were selected empirically, following the method detailed in \citet{AdamsRedai2023}. Images from Epochs 2020 and 2024 were combined into separate primary-subtracted images using inverse-variance weighted averages (Figure~\ref{fig:images}).

The brightness of the star was measured using aperture photometry with a circular aperture of $r=30$\,pixels radius. The position and brightness of the two planets were determined by forward modeling \citep{Pueyo2016}. The PSF representing the planet is created by the HST PSF modeling tool \texttt{Tiny Tim} (native pixel scale=$0.040''$/pixel, \citealt{Krist2011}) and then up-sampled to a pixel scale of $0.020''$/pixel using the same method as applied to the observational data. Forward models were created by the \texttt{pyklip.fm} module. These models were then fit to the PSF-subtracted images and the best-fitting parameters are determined by Markov Chain Monte Carlo (implemented by \texttt{emcee}, \citealt{Foreman-Mackey2013}). The means and standard deviations of the posterior chains were adopted as the best-fitting photometry and astrometry.

\section{Results}
\label{sec:results}
\newcommand{\fluxunit}{\ensuremath{\mathrm{erg\,s^{-1}\,cm^{-2}}}\xspace}

\subsection[The Recovery of PDS 70 b and c]{The Recoveries of \pdsbc}

The primary-subtracted images reveal point sources located at the expected positions of \pdsbc\footnote{The planets' positions are estimated using \url{https://whereistheplanet.com} \citep{Wang2021code}. The orbit solution is provided by \citet{Wang2021}.} (Figure~\ref{fig:images}). We evaluate the detection significance of the two planets using the signal-to-noise ratio (S/N) formula defined in \citep{Mawet2014}. The Epoch 2020 image is consistent with that published by \citet{Zhou2021}. It does not show evidence of \pdsc but detects \pdsb with an S/N of 7. Both planets are present in the Epoch 2024 image. \pdsb is marginally detected with $\mathrm{S/N}=4.0$. \pdsc is more firmly recovered with $\mathrm{S/N} = 5.3$.

The reduction in the detection S/N of \pdsb is due to its decreasing angular separation (see \citet{Wang2021} for the most up-to-date orbit solutions). The background noise at the location of \pdsb is dominated by speckles. Because of \pdsb's Keplerian motion on an inclined orbit, the angular separation of \pdsb decreases from 170\,mas to 149\,mas between the two observing epochs. This results in a significant increase in speckle noise and a more severe impact of ADI over- and self-subtraction in the latter epoch.  Consequently, the detection confidence is lower in the more recent epoch.

\newcommand{\eps}{\ensuremath{\mathrm{e}^{-}/\mathrm{s}}}

\subsection{Astrometry and Photometry}
\label{sec:AstroPhot}

\begin{figure}[!th]
    \centering
    \includegraphics[width=\linewidth]{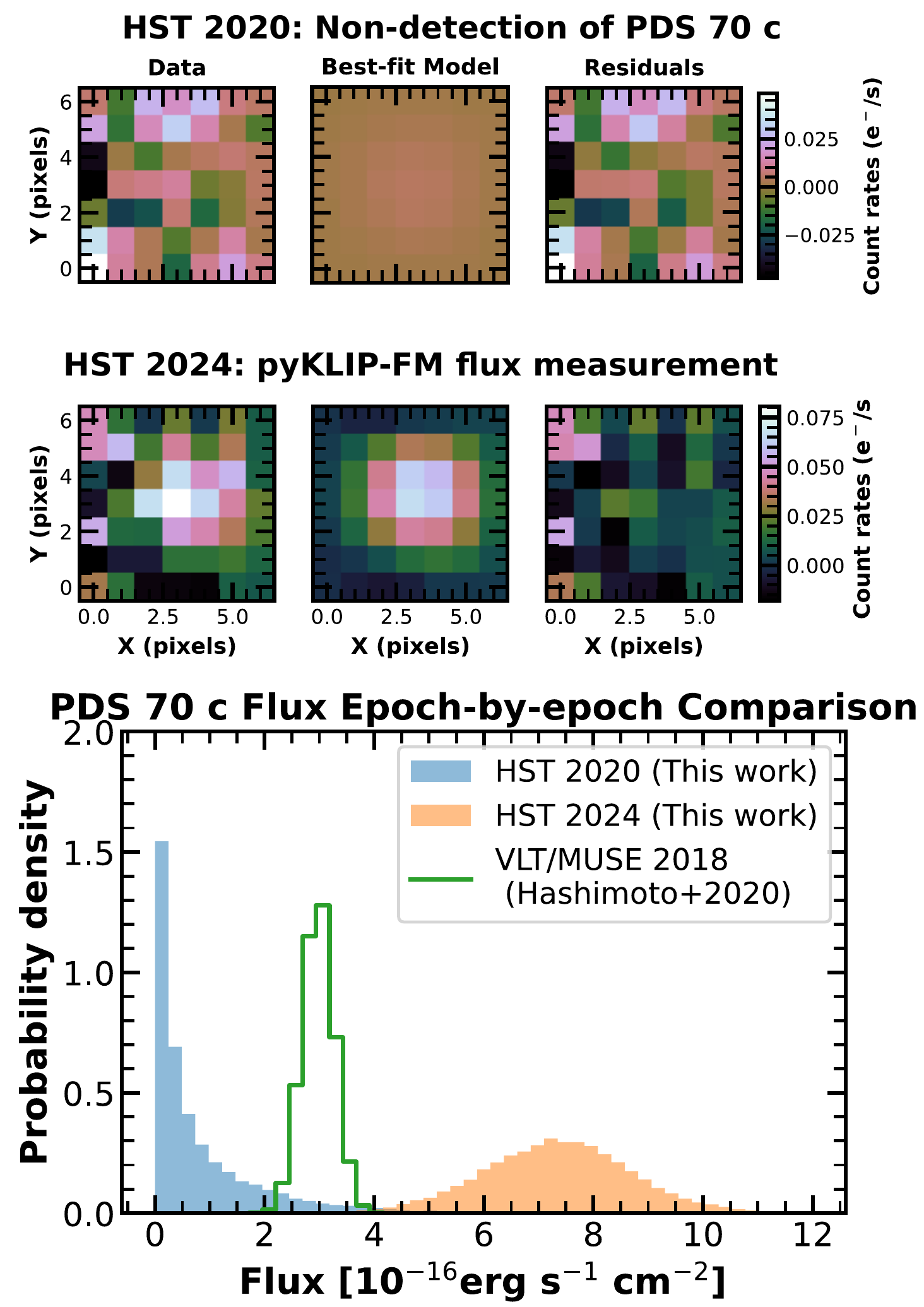}
    \caption{The variable H$\alpha$ flux of \pdsc. Top and middle rows: forward modeling at the expected positions of \pdsc in the 2020 and 2024 data. Both epochs consist of data collected spanning a few months (Table~\ref{tab:obs_data}). The three panels (from left to right) are cut-out data stamps centered at the planet, best-fit forward models, and fitting residuals, respectively. \pdsc is not detected in Epoch 2020, and the model in the top-middle row is consistent with zero flux. Bottom row: a comparison of \pdsc's \halpha{} flux measured in different epochs. Marginal posterior distributions of HST-based measurements are shown by filled histograms (blue and orange). The VLT/MUSE measurement \citep[$f=(3.1\pm0.3)\times10^{-16}\,\fluxunit$;][]{Hashimoto2020} is shown as a green solid line.}
    \label{fig:fluxdistribution}
\end{figure}

The planetary signals are well fit by the forward models of the \texttt{Tiny Tim} point source PSFs and do not show evidence of extended structures. Any self-subtraction between the planet PSFs is corrected by the forward-modeling process \citep{Pueyo2016}. The MCMC fitting implemented by \texttt{pyklip.fm} converges to best-fitting values after the burn-in phase (200 steps). The posteriors of $\Delta$\,RA, $\Delta$\,Dec, and the scaling factors agree with normal distributions. Therefore, the mean and standard deviation values of the posteriors accurately represent the astrometry and photometry of the planets, as well as their associated uncertainties. 

For both planets, the optical continuum flux emitted by their photospheres is negligible compared to that of the H$\alpha$ line. Therefore, we assume all observed flux in the F656N filter is from the emission line. The observed count rate (in \eps) is converted to H$\alpha$ flux by multiplying the conversion factor (\texttt{photflam2} header keyword) and the effective filter width ($W_{\mathrm{eff}}=17.65$\,\AA).  Table\,\ref{tab:result} lists the astrometry and photometry results. Mass accretion rates estimated using model-based scaling laws (Th19: \citealt{Thanathibodee2019} and Ao21: \citealt{Aoyama2021}) are also provided, which range from $\dot{M}\approx1$ to~$3\times10^{-8}\,M_\mathrm{Jup}\,\mathrm{yr^{-1}}$. These estimates are strongly model-dependent and rely on loosely constrained masses and radii. We do not account for potential contamination of the H$\alpha$ flux by stellar light reflected off the circumplanetary disks, which could lead to slight overestimation of the accretion rates.

\begin{deluxetable*}{lcccccccc} 
\tablecaption{Properties of \pdsbc Constrained by the 2024 HST Epoch\label{tab:result}}
\tablehead{
\colhead{Planet} & \colhead{$\Delta$RA} & \colhead{$\Delta$Dec.} & \colhead{PA} & \colhead{Sep} & \colhead{Count Rate} & \colhead{F656N Flux} & \colhead{$\log\dot{M}_\mathrm{Th19}$} & \colhead{$\log\dot{M}_\mathrm{Ao21}$}\tablenotemark{b} \\
\colhead{} & \colhead{(mas)} & \colhead{(mas)} & \colhead{($^\circ$)} & \colhead{(mas)} & \colhead{(\eps)} & \colhead{($10^{-16}\,\mathrm{erg}\,\mathrm{s^{-1}}\,\mathrm{cm^{-1}}$)}
&\colhead{$M_\mathrm{Jup}\,\mathrm{yr^{-1}}$} &\colhead{$M_\mathrm{Jup}\,\mathrm{yr^{-1}}$}
}
\startdata
\pdsb & $110\pm28$ & $-97\pm24$ & $131.4\pm5.6$ & $147\pm28$ & $1.30\pm1.24$\tablenotemark{a} & $4.0\pm3.7$\tablenotemark{a} & $-8.0$ & $-7.5$ \\
\pdsc & $-213\pm7$ & $-2.7\pm6.5$ & $213.5\pm7.0$ & $269.2\pm1.7$ & \fluxcepsII & \fluxcflamII & $-8.0$ & $-7.6$\\
\enddata
\tablenotetext{a}{The flux uncertainties of \pdsb are strongly impacted by the fitting uncertainties in its positions and surrounding speckles. S/N values reported in Section~\ref{sec:AstroPhot} are calculated based on predetermined planet positions. Consequently, $f_\mathrm{b}/\Delta f_\mathrm{b} < \mathrm{S/N}_\mathrm{b}$.} 
\tablenotetext{b}{We adopt planetary masses and radii of $M_\mathrm{b}=3.2\,M_\mathrm{Jup}$, $M_\mathrm{c}=7.5\,M_\mathrm{Jup}$, $R_\mathrm{b}=2.0\,R_\mathrm{Jup}$, and $R_\mathrm{c}=2.0\,R_\mathrm{Jup}$ \citep{Wang2021, Blakely2024}.}
\end{deluxetable*}

\subsection[Evidence for Variable Planetary H alpha Flux]{Evidence for Variable Planetary H$\alpha$ Flux}\label{sec:evidence}

\begin{figure}[t]
    \centering
    \includegraphics[width=\linewidth]{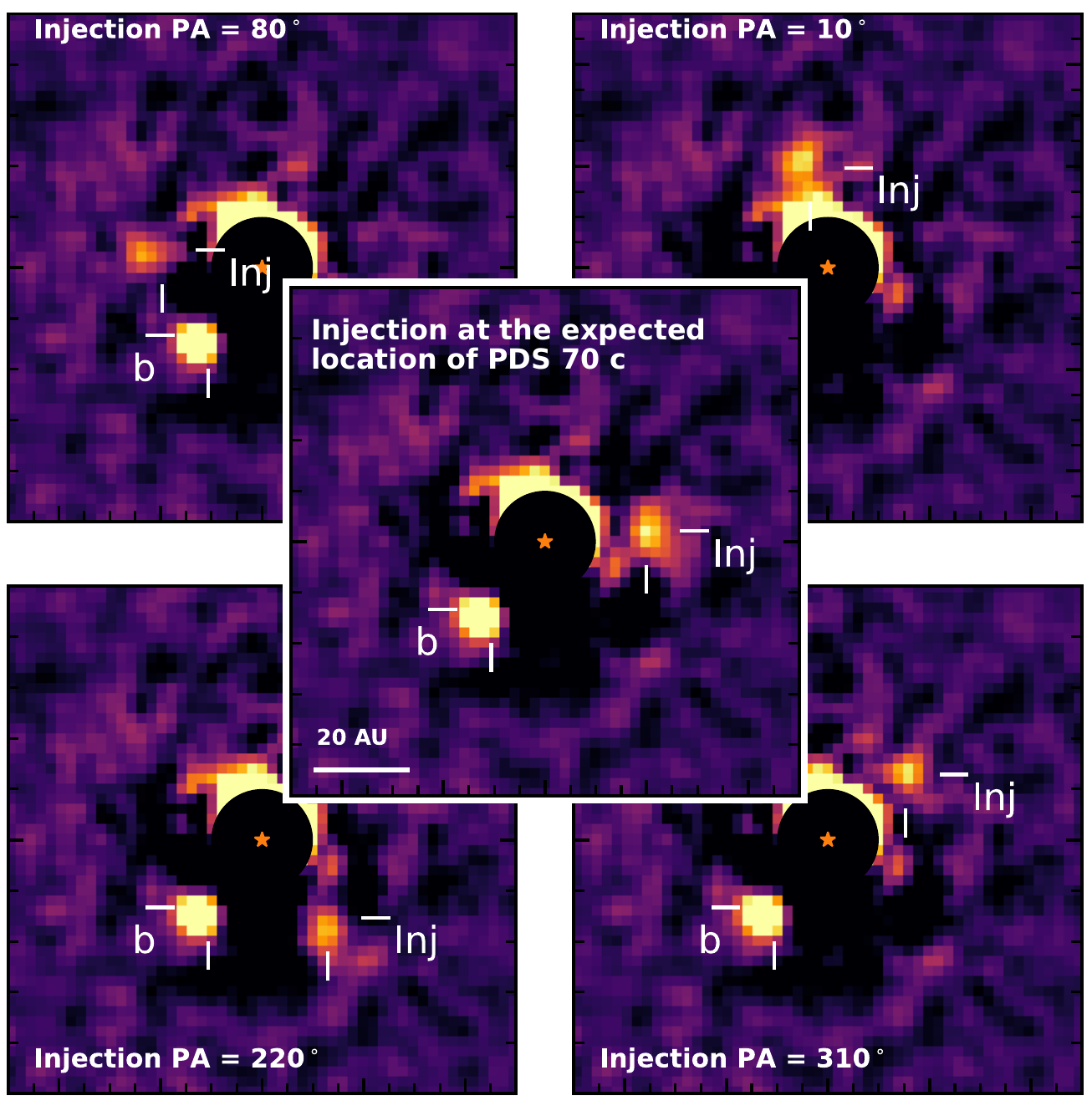}
    \caption{H$\alpha$ variability of \pdsc{} confirmed by injection/recovery tests. Point sources that have the same flux as \pdsc measured in Epoch 2024 are injected into the Epoch 2020 frames at the expected position of \pdsc (middle) and four different position angles. The injected planets are confidently recovered in all cases.}
    \label{fig:injection}
\end{figure}

The recovered H$\alpha$ flux of \pdsbc differs from those reported in previous studies \citep{Haffert2019,Hashimoto2020,Zhou2021}. The 2024 flux of \pdsb ($f_\mathrm{b,\,2024}=(4.0\pm3.7)\times10^{-16}\,\fluxunit$) is less than 1/4 of the HST 2020 epoch flux ($f_\mathrm{b,\,2020}=(1.65\pm0.2)\times10^{-15}\,\fluxunit$). The 2024 flux of \pdsc ($f_\mathrm{c,\,2024}=(\fluxcflamII)\times10^{-16}\,\fluxunit$) is more than twice its H$\alpha$ flux ($f_\mathrm{c,\,MUSE}=(3.1\pm0.3)\times10^{-16}\,\mathrm{erg}\,\mathrm{s^{-1}}\,\mathrm{cm^{-1}}$) measured using VLT/MUSE data taken in 2018 (Figure~\ref{fig:fluxdistribution}). Variations in flux measurements between space- and ground-based observations may arise from the discrepancies in photometric calibration methods.

Forward-modeling photometry constrains the significance and amplitude of the variability signals. The H$\alpha$ flux of \pdsb  is not accurate enough to confirm its variability due to its close separation to the star in Epoch 2024 and the impact of self-subtraction. The variability of \pdsc can be precisely quantified. Figure~\ref{fig:fluxdistribution} shows the marginal distribution of \pdsc flux constrained by the 2020 and 2024 data. The Epoch 2024 \halpha{} flux of \pdsc ($f_\mathrm{c,\,2024}=(\fluxcflamII)\times10^{-16}\,\fluxunit$) is ${>}3\sigma$ greater than the upper limit of the Epoch 2020 flux, defined as the 95 percentile of its posterior distribution ($f_\mathrm{c,\,2020}<\fluxcflamI\times10^{-16}\,\fluxunit$).  A 100\% or greater increase in H$\alpha$ flux is required to explain the difference between the two epoch images.

Injection-and-recovery tests further verify the variability signal. We inject synthetic PSFs that best fit to \pdsc in Epoch 2024 to the Epoch 2020 data. The injection location is first conducted at the expected position of \pdsc ($\mathrm{PA}=276^\circ$, $\mathrm{separation}=215\,\mathrm{mas}$) and then repeated at four other locations with different PAs ($10^\circ$, $80^\circ$, $220^\circ$, and $310^\circ$) and an identical separation (215\,mas). All four injected signals are recovered with high confidence ($\mathrm{S/N}>3$, Figure~\ref{fig:injection}). If \pdsc had a constant brightness, it should have been detected in the 2020 epoch. 

Several of our data reductions show evidence for variability in \pdsb between 2020 and 2024, as well as in \pdsc between 2024-April and 2024-May, however, these variability signals strongly rely on the choice of \texttt{pyKLIP} control parameters. These low-significance variability signals are susceptible to over- or under-subtraction of stellar speckles \citep{Pueyo2016}. For \pdsb, the uncertainties originates from the strong speckle noise at a close angular separation to the star. In the case of \pdsc{}, the detection of month-to-month variability is limited by the restricted variation in telescope orientation within each visit-set. Currently, only the 2020 vs. 2024 variability in \pdsc is robustly confirmed by our data. We defer the confirmation of the variability in \pdsb and short-cadence variability in \pdsc to future studies after the complete dataset is collected in 2025 Spring.

\subsection[Variable H alpha Flux of the Host Star]{Variable H$\alpha$ Flux of the Host Star}

\begin{figure}[t]
    \centering
    \includegraphics[width=\linewidth]{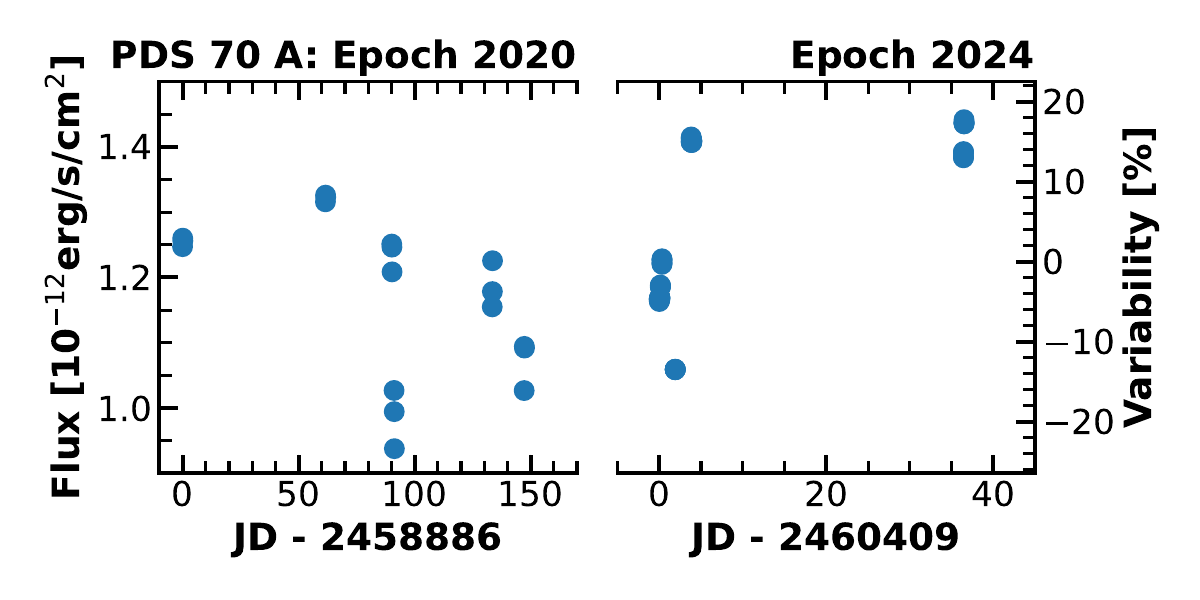}
    \caption{Variable H$\alpha$ flux emitted by the host star. Flux measured in Epoch 2020 and Epoch 2024 are shown as orange squares (left) and blue circles (right), respectively. Uncertainties are negligible ($<1\%$) compared to the variability. Scales of absolute flux and relative changes are shown on the left and right axes, respectively. The variability is defined by the relative change over the median flux value.}
    \label{fig:star}
\end{figure}

The star PDS\,70 also has variable H$\alpha$ flux  (Figure~\ref{fig:star}). The peak-to-peak flux change is over 40\%, higher than that observed in the broad visible band \citep{Thanathibodee2020}. In Epoch 2024, the mean and standard deviation of the H$\alpha$ flux are $f=1.27\times10^{-12}$ and $1.31\times10^{-13}$\,\fluxunit, respectively. The Epoch 2024 average flux is higher than that measured in Epoch 2020 ($f=1.18\times10^{-12}$\,\fluxunit), but the difference is less than $1\sigma$.

\subsection[The Outer Disk of PDS 70 in Scattered Light]{The Outer Disk of PDS\,70 in Scattered Light}


The Epoch 2024 image also detected the disk scattered light. We estimate the location of the disk scattered light and the possible contamination to the planet flux using the \texttt{diskfm} module of the \texttt{pyKLIP} package. For simplicity, a geometric disk model, following \citet{Blakely2024}, is adopted over a radiative transfer model. The disk position angle and inclination are fixed to $160^\circ$ and $53.9^\circ$ based on prior constraints \citep[e.g.,][]{Benisty2021,Blakely2024}. We construct nine models to sample the peak de-projected radial location from $0.34''$ to $0.50''$. Each model comprises an axisymmetric component and an asymmetric component accounting for differences in scattering angle. The scalings of the model components are optimized to minimize the sum squared residuals of pixels with de-projected distance between 40 to 80 AU ($0.35''$ to $0.70''$) away from the star. Two circular regions ($r=0.04''$) centered at the planets  are excluded from the residual calculations. The scattering light brightness is assumed to be constant over time.

\begin{figure}[!t]
    \centering
    \includegraphics[width=\columnwidth]{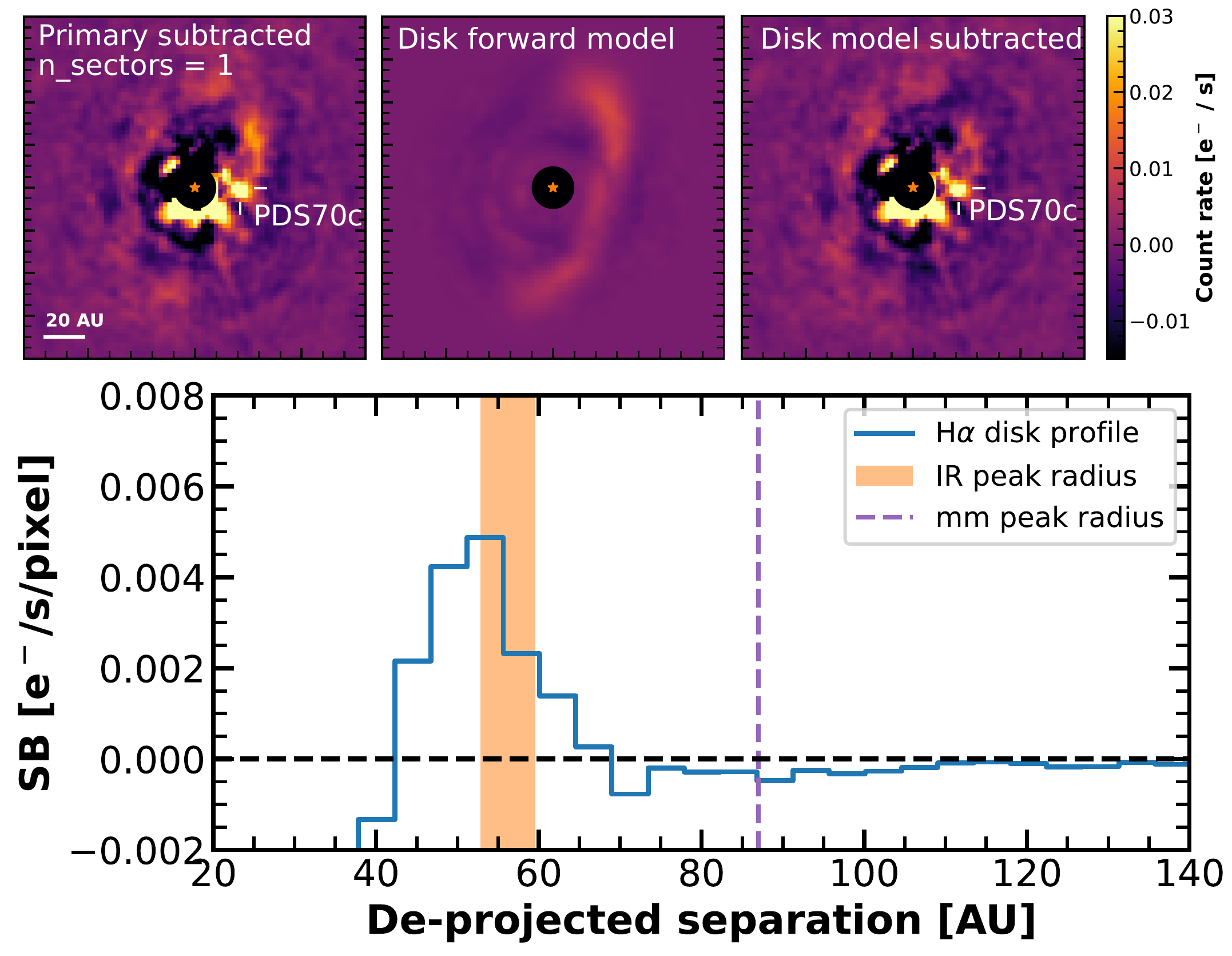}
    \caption{Estimations of the disk scattered light emission (top panels) and the disk radial profiles (bottom panel). The top panels include a primary-subtracted image (left), the best-fitting model (middle), and the disk subtraction residuals. The primary subtraction was conducted using a relatively less aggressive strategy ($n_\mathrm{sector}=1$, i.e., no sub-division within an annulus) to preserve disk flux. The H$\alpha$ scattered light of the disk can be reasonably well fit by a simple geometric model. In the radial direction, the disk's H$\alpha$ scattered light peaks at $50\,\mathrm{au}$, slightly closer to the star than the disk seen in infrared. The disk surface brightness is low and has negligible contaminations to the planets' flux.} \label{fig:disk}
\end{figure}

Figure~\ref{fig:disk} presents the disk forward modeling results. A radius of $r=0.42''$ disk minimizes the residuals. However, the disk flux is not completely eliminated by subtracting the best-fitting disk model, suggesting more complex disk structures than the geometric model. The radial peak is near the inner edge of continuum dust emission \citep[e.g.,][]{Benisty2021} and slightly closer to the star than the one best-fitting to the infrared data \citep{Blakely2024}, suggesting micron-sized dust filtering into the transition disk gaps and trapped in pressure bumps \citep[e.g.,][]{Pinilla2018}. The disk surface brightness measured at the positions of \pdsbc is less than 0.002 e$^-$\,s$^{-1}$. The disk contamination contributes less than 5\% of the observed flux of  \pdsbc, negligible compared to the measurement uncertainties.

\section{Discussion}
\label{sec:discussion}
\newcommand{\LHA}{\ensuremath{L_{\mathrm{\halpha}}}\xspace}
\newcommand{\Lacc}{\ensuremath{L_{\mathrm{acc}}}\xspace}

\subsection[A Comparison with MagAO-X Results]{A Comparison with the MagAO-X Results}

The \halpha variability trends of \pdsbc are consistently observed in both HST (this work) and MagAO-X \citep{Close2025} data. Both datasets concur that \pdsb has dimmed, \pdsc has brightened, and \pdsc surpassed \pdsb in \halpha{} luminosity in 2024 March/April. It is important to note that the HST \pdsb measurement in the 2024 Epoch is significantly affected by speckle noise and should be interpreted with caution. 

The observed \halpha brightening in \pdsc shows coincidentally similar amounts between this work ($2.4\times$, between 2020 Spring and 2024 Spring) and the MagAO-X results ($2.3\times$, between 2023 March and 2024 March). The four visit-sets of the 2024 HST observations (2024-04-08, 2024-04-10, 2024-04-12, and 2024-05-15) happened two to seven weeks after the MagAO-X  observation on 2024-03-25. The HST line flux exceeds the MagAO-X flux by 50\% with a$1.8\sigma$ significance ($f_\mathrm{HST}=(\fluxcflamII)\times10^{-16}\,\fluxunit$, $f_\mathrm{MagAO{-}X}=(4.78\pm0.46)\times10^{-16}\,\fluxunit$). This provides tentative evidence supporting short-term variability in \pdsc between March and May 2024. However, systematic uncertainties due to differences in photometric calibrations between space and ground-based observations cannot be ruled out as the origin of the flux discrepancy. 

\citet{Close2025} reported the detection of optical continuum flux at the position of \pdsb and attributed it to reflected stellar light from a circumplanetary disk. They suggested that this flux could contaminate the \halpha line flux and applied a correction factor of $1/1.54$  to the HST Epoch 2020 flux when comparing the HST and MagAO-X photometry of \pdsb. Although no significant visible continuum flux has been detected near \pdsc, similar reflected light from \pdsc's disk might be obscured by speckle noise. Applying the same correction factor to our HST flux brings it consistent with the 2024 MagAO-X result.

\subsection[Possible Origins of H alpha Variability]{Possible Origins of H$\alpha$ Variability}


PDS\,70\,c's H$\alpha$ luminosity variability is within the range of that observed over year-long timescales in the Pa$\beta$ emission of the substellar companions GQ~Lup~b and GSC~06214~b (10--$30~\MJ$) as studied by \citet{Demars2023}. The spectroscopically resolved hydrogen recombination lines observed at these objects provide evidence that they grow similarly to classical T Tauri stars by magnetospheric accretion \citep[e.g.,][]{Ringqvist2023, Viswanath2024}. T Tauri stars are variable over a wide range of timescales in the form of rotational modulations and various stochastic patterns \citep[e.g.,][]{Costigan2014,Zsidi2022,Herczeg2023}. If the variability observed in \pdsc has the same origin as the T Tauri stars, it must be in the long-term stochastic form because the observed variability timescale (a few years) is too long for rotational modulation. Indeed, the rotation period is expected to range from 10 to 20\,hr down to less than an hour, nearly the break-up limit (see, e.g., \citealt{Bryan2018, Hasegawa2021}). Then, the possible cause of the variability includes unstable interactions between the planet's magnetosphere and the circumplanetary disk \citep[similar to those simulated for stars, see e.g.,][]{Kulkarni2008,Takasao2022} or density fluctuations within the circumplanetary disk, whose Keplerian orbital timescale is less than 20 years \citep[estimated based on the disk size;][]{Benisty2021}.

Hydrogen line emission from the post-shock gas, as opposed to the heated pre-shock gas in magnetospheric accretion, is also proposed as a scenario to model the accreting planets' H$\alpha$ emission \citep{Aoyama2018, Aoyama2019}. A hydrodynamic model that couples accretion onto the disk and the planet suggests that the accretion shock at the planet's surface dominates the H$\alpha$ emission and that the accretion onto the disk triggers high-frequency (timescale $\ll10$\,days), stochastic, and high-amplitude (order-of-magnitude) variability \citep{Takasao2021}. The detailed predictions could be sensitive to the details of the thermodynamics and the radiative transfer \citep{Marleau2023}. The modeled H$\alpha$ luminosity smooths out over long timescales ($>10$\,days). Testing the feasibility of the postshock gas emission in reproducing H$\alpha$ variability observed on year-long timescales requires extending the time length of existing models \citep[e.g., ][]{Takasao2021}.

Recently, \citet{Christiaens2024} identified a spiral-arm-like feature connecting \pdsc and the outer transition disk and interpreted it as a stream feeding the planet's disk. These findings draw parallels between accreting planets and protostars fed by streamers, in which interaction between the disk and the streamer flow triggers accretion outbursts \citep[e.g.,][]{Pineda2020}. JWST/NIRISS/AMI and NIRCam observations separated 12 days apart found a 75\% (${\sim}3\sigma$) discrepancy in the F480W flux of \pdsc \citep{Blakely2024}. If this difference is confirmed to trace variability of the circumplanetary disk (as against to instrumental or data-reduction-related systematic errors), this points to instability in the disk that may lead to the variability of accretion onto the planet.

Variations in line-of-sight extinction, potentially caused by micron-sized dust in the circumplanetary disk or by shielding from the accretion flow, can also drive the variability in \halpha{} flux \citep[e.g.,][]{Szulagyi2020,Marleau2022}. A decrease of $\Delta A_V>0.9$\,mag in extinction can account for the observed $>2\times$ H$\alpha$ flux increase from 2020 to 2024. Simultaneous observations of accretion tracers in the visible and infrared bandpasses, even in the cases of non-detections, help constrain extinction level and test whether changing extinction is the source of H$\alpha$ variability \citep{Hashimoto2020, Uyama2021}.

By contributing to the planet's visible band flux through circumplanetary disk scattering, the stellar light may impact the observed variability of the planets (\citealt{Close2025}; see also a light-echo experiment with the AB Aurigae system, Bowler et al., submitted). Variability of the host star between the 2020 and 2024 epochs (the epoch-to-epoch average flux difference is $<10\%$) is too low to explain the siginificant variability observed in \pdsc. Future observations with improved cadence and sensitivity may distinguish the host star and planetary accretion-induced variability by identifying correlated changes between the planetary and stellar variability (Bowler et al.\ submitted).

\subsection[Enabling an Empirical Lacc--LHalpha Calibration for Accreting Planets]{Enabling an Empirical \Lacc--\LHA Calibration for Accreting Planets}

\begin{figure}[t]
    \centering
    \includegraphics[width=\linewidth]{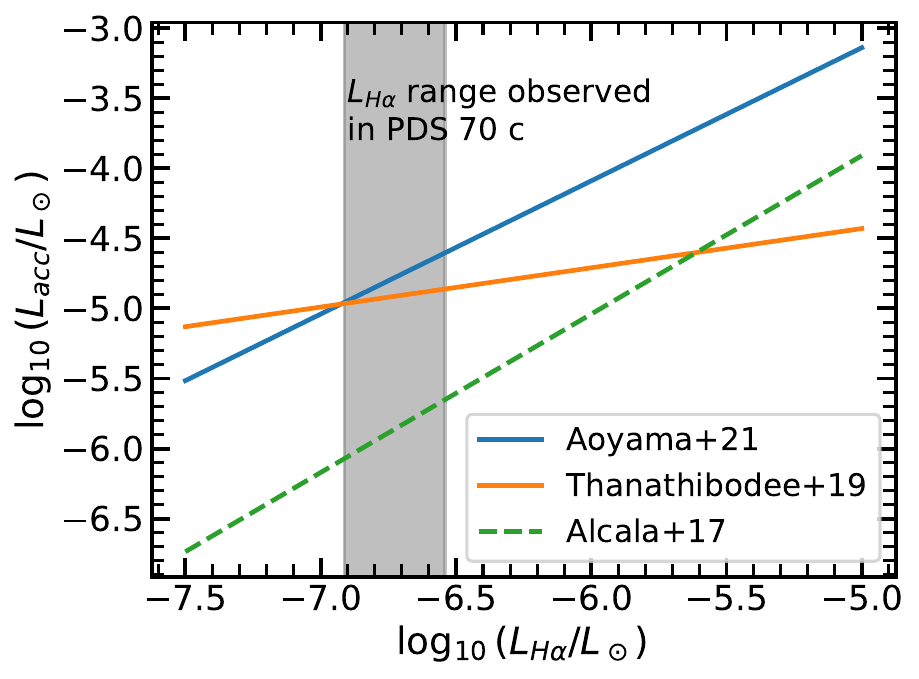}
    \caption{Comparisons of the \Lacc–\LHA relationships near the observed \LHA of \pdsc. Predictions from the planetary accretion shock model (\citealt{Aoyama2021}, blue solid line), the planetary magnetospheric model (\citealt{Thanathibodee2019}, orange solid line),  and the empirical relation observed in T Tauri stars (\citealt{Alcala2017}, green dashed line) are shown. The vertical grey region highlights the range of \LHA observed in \pdsc. Empirically mapping the \Lacc–\LHA relationship in the PDS 70 planets will test these predictions.}
    \label{fig:expected_variability}
\end{figure}

An important caveat is that these interpretations all rely on a correlation between the H$\alpha$ luminosity and the mass accretion rate. The H$\alpha$ emission from accreting stars often contain chromospheric components. The long-term monitorings of accreting low-mass stars show that the H$\alpha$ luminosity poorly constrains the accretion rate \citep{Herczeg2023}. The large-amplitude H$\alpha$ variability observed in \pdsc provides a new avenue to test the validity of $\LHA$ in estimating the accretion rate if multi-epoch H$\alpha$ and direct \Lacc measurements can be obtained. Testing the \LHA--\Lacc on one planet avoids the inherent physical differences compared to other objects and isolates the origin of the H$\alpha$ emission and its luminosity variability.

Figure~\ref{fig:expected_variability} shows the expected accretion luminosity variations with the changing H$\alpha$ luminosity. The assumed \LHA--\Lacc relations, which include the prediction from the planetary strong shock model \citep{Aoyama2018,Aoyama2021}, planetary magnetospheric model \citep{Thanathibodee2019}, and the empirical classic T Tauri star relation \citep{Alcala2017}, differ significantly. Simultaneously monitoring \pdsbc in multiple accretion tracers, which has been carried out for low-mass stars and substellar companions \citep[e.g.,][]{Betti2022,Aoyama2024,Hashimoto2025}, will allow for a robust estimate of accretion luminosity variability. Then, the \LHA--\Lacc relation observed in \pdsbc informs their mass accretion mechanisms.

\subsection{Implications about Detecting Accreting Planets}

Variability of the H$\alpha$ emission is proposed as a possible solution to explain the low yield of accreting planet detections \citep[e.g..][]{Brittain2020}. Our observations of \pdsc verify that the high-amplitude variability can critically impact the detectability of the planet. However, variability alone does not fully explain the lack of similar discoveries to \pdsbc. Ground-based high-contrast H$\alpha$ imaging surveys have conducted over 50 unique observations on over 15 transition disk systems \citep{Zurlo2020,Xie2020,Follette2023}. Assuming all systems have at least one \pdsc-like planet and all planets have an identical accretion duty cycle, the accreting epoch must be very short such that the possibility of catching the planet while they are accreting is close to zero. This is incompatible with the successful H$\alpha$ recoveries of \pdsbc in multiple epochs \citep[][and this work]{Wagner2018,Haffert2019,Zhou2021}. It is much more likely that PDS\,70 bears unique disk and planetary properties that significantly enhance the direct detectability of the planets compared to other transition disk targets. Continued monitoring of \pdsc will help constrain the timescales and duty cycle of H$\alpha$ bursts, which is critical for accurately assessing imaging survey yields and estimating the occurrence rates and demographics of young gas giants.

\section{Summary}
\label{sec:summary}
Our main findings are the following:
\begin{enumerate}
\item We obtained new HST/WFC3/UVIS \halpha{} narrowband imaging data of the PDS\,70 system and combined these data with a previous HST epoch taken in 2020. The new images have four times longer integration time than the previous epoch, providing a more sensitive view of this site of planet formation. \pdsc and scattered light of the outer disk of PDS\,70 are detected for the first time in HST \halpha{} images.

\item Planets \pdsb and \pdsc were recovered in the new images with S/Ns of 4.0 and 5.3, respectively. The positions of the two planets are consistent with those predicted by the latest orbit fitting solutions within their respective $1\sigma$ uncertainties. 

\item Assuming all of the observed flux comes from the \halpha line emission, the \halpha line fluxes are $f=(4.0\pm3.7)\times10^{-16}\,\fluxunit$ and $f=(\fluxcflamII) \times 10^{-16}\,\fluxunit$ for b and c, respectively. The corresponding mass accretion rates are $\dot{M}\approx1$ to~$3\times10^{-8}\,M_\mathrm{Jup}\,\mathrm{yr^{-1}}$ based on scaling laws derived from planetary accretion models \citep{Thanathibodee2019, Aoyama2021}.

\item The \halpha{} flux of \pdsc shows significant variability between the 2020 and 2024 epoch. The Epoch 2024 flux is 2.4 times the upper limit ($95$ posterior percentile) obtained from the Epoch 2020 data ($>3\sigma$ confidence). The 2024 flux is also more than double the H$\alpha$ flux measured with 2018 VLT/MUSE data. Injection-and-recovery tests confirm the variability of \pdsc between the two HST epochs.

\item The \halpha{} flux of \pdsb measured in the 2024 epoch is less than 25\% of its epoch 2020 level reported by \citet{Zhou2021}. While this aligns with the the planet's decreasing \halpha{} luminosity determined using the MagAO-X data \citep{Close2025}, the strong speckle noise at \pdsb's 2024 epoch location in the HST data prevents us from definitively confirming the variability in \pdsb.

\item The variability trends observed with HST are consistent with those observed with ground-based MagAO-X instrument \citep{Close2025}. \pdsc's Epoch 2024 HST flux is ${\sim}50\%$ greater than that observed with MagAO-X a few weeks earlier, which tentatively indicate short-term variability. Confirming this variability requires calibrating space-based and ground-based high-contrast imaging photometry.

\item The non-detection of \pdsc in the 2020 epoch and its recovery in the 2024 epoch provides a concrete example where variable \halpha flux can impact the yields of protoplanetary imaging survey in the \halpha bandpass. However, \halpha variability is unlikely to be the sole reason for the low yields of \mbox{(proto-)}planet searches. Future monitoring observations are necessary to determine the periodicity and duty cycle of the accretion activity. Estimating the yield of the imaging survey results should account for the variability flux to deliver unbiased population statistics of protoplanets. 
\end{enumerate}

\software{Astropy \citep{astropy:2013,astropy:2018}, Scipy \citep{scipy:2020}, Numpy \citep{numpy:2020}, Matplotlib \citep{matplotlib:2007}, Seaborn \citep{Waskom2021}, pyKLIP \citep{pyklip:2015}, emcee \citep{Foreman-Mackey2013}}

\vspace{1em}
We acknownlege the anonymous referee for a prompt and constructive report that improved the manuscript. This work would not be possible without the excellent observation support from STScI Program Coordinator Amber Armstrong and Contact Scientist Joel Greene.
This research is based on observations made with the NASA/ESA Hubble Space Telescope obtained from the Space Telescope Science Institute, which is operated by the Association of Universities for Research in Astronomy, Inc., under NASA contract NAS 5–26555. These observations are associated with programs GO-15830, GO-16651, and GO-17427. Y.Z.\ acknowledges HST data analysis grants associated with programs GO-15830, GO-16651, and GO-17427. This material is based upon work supported by the National Science Foundation Graduate Research Fellowship under Grant No.~2139433. G.-D.M.\ acknowledges the support from the European Research Council (ERC) under the Horizon 2020 Framework Program via the ERC Advanced Grant ``Origins'' (PI: Henning), Nr.~832428,
and via the research and innovation programme ``PROTOPLANETS'', grant agreement Nr.~101002188 (PI: Benisty). S.T.\ was supported by JSPS KAKENHI grant Nrs.~JP21H04487, JP22K14074, and~JP22KK0043. J.H.\ acknowledges the support from JSPS KAKENHI Grant Nr.~23K03463. Y.H.\ was supported by the Jet Propulsion Laboratory, California Institute of Technology, under a contract with the National Aeronautics and Space Administration (80NM0018D0004).

\appendix
\section[Summary of HST High-Contrast Observations of PDS 70]{Summary of HST High-Contrast Observations of PDS\,70}
\label{sec:appendix}
Information of HST observations of PDS\,70 is summarized in Table\,\ref{tab:obs_data}. All observations are taken with WFC3/UVIS2 paired with the F656N filter. All data used in this paper can be accessed at \url{http://dx.doi.org/10.17909/ndk6-9441}.

\begin{deluxetable*}{ccccccc}
\tablecaption{Observational Data Summary\label{tab:obs_data}}
\tablewidth{0pt}
\tablehead{
    \colhead{Visit-Set} & 
    \colhead{Beginning Time (UT)} & 
    \colhead{End Time (UT)} & 
    \colhead{Orbits} & 
    \colhead{Position Angles ($^\circ$)} & 
    \colhead{$N_{\rm exp}$} & 
    \colhead{Total Exp.\ Time (s)}
}
\startdata
\hline
\multicolumn{7}{c}{Epoch 2020, PID: 15830}\\\hline
1 & 2020-02-07 00:40:16& 2020-02-07 04:29:40 & 3 & 100.8, 110.8, 120.8 & 27 & 540 \\
2 & 2020-04-08 14:27:06& 2020-04-08 18:19:53 & 3 & 140.8, 152.0, 163.0 & 27 & 540 \\
3 & 2020-05-07 04:41:56& 2020-05-07 08:33:35 & 3 & 181.0, 191.0, 201.0 & 27 & 540 \\
4 & 2020-05-08 04:31:21& 2020-05-08 08:23:00 & 3 & 211.0, 221.0, 231.0 & 27 & 540 \\
5 & 2020-06-19 11:52:50& 2020-06-19 15:44:29 & 3 & 259.1, 269.1, 279.1 & 27 & 540 \\
6 & 2020-07-03 06:19:48& 2020-07-03 10:06:19 & 3 & 279.1, 289.1, 297.1 & 27 & 540 \\
\hline
\multicolumn{7}{c}{Epoch 2024, PID: 17427}\\\hline
1 & 2024-04-08 19:12:26 & 2024-04-09 03:40:00 & 6 & 135.3, 155.4 & 144 & 8448\\
2 & 2024-04-10 17:04:57 & 2024-04-10 19:23:33 & 2 & 136.9, 146.9 & 48 & 2880\\
3 & 2024-04-12 14:56:59 & 2024-04-12 17:15:34 & 2 & 139.6, 154.6 & 48 & 3000\\
4 & 2024-05-15 05:18:34 & 2024-05-15 07:38:59 & 2 & 216.6, 230.6 & 48 & 3120\\
\hline
\enddata

\end{deluxetable*}

\bibliographystyle{aasjournal.bst}
\bibliography{references.bib}

\end{document}